\documentclass{aa}

\usepackage{graphicx}
\usepackage{txfonts}
\usepackage{natbib}
\bibpunct{(}{)}{;}{a}{}{,} % to follow the A&A style

\usepackage[normalem]{ulem}
\usepackage{color}
\definecolor{revisionthree}{rgb}{0,0,1}

\usepackage{url}

\begin{document}

   \title{Rotation and outburst-like brightenings in ZZ~Ceti stars observed by TESS
   }

   %\subtitle{}

    \author{Zs\'ofia~Bogn\'ar\inst{}\thanks{\email{bognar.zsofia@csfk.org}},
         \'Ad\'am~S\'odor\inst{}}
   
   \institute{
        Konkoly Observatory, HUN-REN Research Centre for Astronomy and Earth Sciences, MTA Centre of Excellence, Konkoly Thege Mikl\'os \'ut 15-17, 1121 Budapest, Hungary
        }
        
    \titlerunning{Rotation and outbursts in ZZ~Ceti stars}
        \authorrunning{Zs.~Bogn\'ar et al.}
        
    \date{}

% \abstract{}{}{}{}{} 
% 5 {} token are mandatory
 
  \abstract
   {Pulsating white dwarfs, such as ZZ~Ceti stars, allow for the study of white dwarf rotation through rotationally split pulsation modes. Space-based photometry has also revealed outbursts in some cool ZZ~Ceti stars.
   }
    {We extend the search for rotationally split pulsation multiplets and outburst-like brightenings among known ZZ~Ceti stars observed by TESS. We check whether the TESS data add new evidence for a connection between outburst activity and stellar rotation.
    }
     {We analysed 120 s and 20 s cadence TESS light curves available up to Sector 96. After light curve correction and quality filtering, we performed iterative frequency analysis. Candidate rotationally split multiplets were identified automatically and visually validated. Outburst candidates were selected from smoothed flux excursions, followed by light curve inspection and pixel-level TESS diagnostics.}
   {We identify 11 ZZ~Ceti stars with accepted triplet candidates and derive rotation periods of about 8--94~h; three triplet detections are reported here for the first time. The targets are broadly consistent with the dispersion of rotation periods found in previous studies. The time-domain search produced 19 outburst-like events in 11 stars after light curve inspection. Pixel-level diagnostics indicate that most are likely caused by foreground asteroid contamination. We identify three stars (TIC~219442838, TIC~307908863, and TIC~630426257) as candidate stellar outbursters. None belong to the accepted triplet-rotation sample.
   }
   {}


   \maketitle

   \nolinenumbers
%
%________________________________________________________________

\section{Introduction}

ZZ~Ceti (or DAV) stars constitute the most populous class of pulsating white dwarfs. These low-to-mid radial-order $g$-mode pulsators show brightness variations in the 100--1500\,s period range, with observed amplitudes spanning from the sub-millimagnitude level in space-based photometry to percent-level variations in high-amplitude objects \citep[see the review by][]{2019A&ARv..27....7C}. The $\kappa-\gamma$ mechanism \citep{1981A&A...102..375D, 1982ApJ...252L..65W} in combination with the convective driving mechanism \citep{1991MNRAS.251..673B, 1999ApJ...511..904G} is responsible for the excitation of the observed pulsations. The recent statistical study by \citet{2025JApA...46...81C} compiled stellar parameters and oscillation properties for 339 known ZZ~Ceti stars, 194 of which had been analysed asteroseismologically.

Their behaviour can be well understood in terms of their location within the instability strip during their cooling evolution \citep[e.g.][]{2008PASP..120.1043F,2010A&ARv..18..471A,2019A&ARv..27....7C}. When entering the hot, blue edge of the empirical instability strip, they generally exhibit only a few low-amplitude brightness variations with short pulsation periods. Such short-period modes are often the most favourable ones for precise pulsation timing and, in selected cases, for detecting evolutionary, secular period changes; see, e.g., \citet{2021ApJ...906....7K}. As the star cools, however, the observed amplitudes increase, and more modes become detectable. This temperature--period progression is also reflected in statistical samples of DAVs, including the weighted-mean-period analysis of \citet{2025JApA...46...81C} and the Kepler/K2 sample discussed by \citet{2017ApJS..232...23H}. 
Amplitude and frequency variability have been reported in individual pulsating white dwarfs, and detailed continuous space photometry has demonstrated such behaviour particularly clearly for the DBV star KIC~08626021 \citep{2016A&A...585A..22Z}. For the present analysis, the relevant practical point is that richer DAV pulsation spectra, unresolved close peaks, and time-dependent amplitudes can complicate the determination of pulsation frequencies using standard pre-whitening processes. The impact of continuous space-based monitoring on the study of pulsating white dwarfs is reviewed by \citet{2020FrASS...7...47C}.

Two phenomena that are discussed later in this paper should be mentioned here. The first is rotational frequency splitting. Rotation lifts the azimuthal degeneracy of non-radial modes, producing multiplets whose frequency separations depend on the rotation rate in the layers probed by the corresponding pulsation modes. In pulsating white dwarfs, we can usually observe only modes with horizontal degrees $\ell=1$ and $2$, whose multiplets appear as triplets and quintuplets, respectively. We focus on triplets because a complete $\ell=1$ pattern provides the cleanest observational case for measuring a single rotational splitting: the two side components should be placed approximately symmetrically around the central component, while incomplete doublets or more complex multiplets require additional assumptions about the missing azimuthal components or the mode degree. In this paper, we search for triplet frequencies to determine the stellar rotation periods. That is, we continued our work presented first in \citet{2024A&A...684A..76B}, hereafter B24, but with more data sets and a more systematic splitting-search pipeline. Instead of inspecting Lomb-Scargle periodograms by eye as the primary search method, we first generated pre-whitened frequency lists for all light curves, automatically searched them for symmetric triplet patterns, grouped candidates with similar splittings into consensus clusters, ranked the clusters using internal quality metrics, and only then performed the final visual validation.

More broadly, the motivation for searching for multiplets is that white dwarf rotation rates, especially as a function of stellar mass, can provide insight into the angular-momentum transport mechanisms operating in their progenitors, including the coupling between red-giant cores and envelopes \citep{2017ApJS..232...23H}. Further discussion can be found in \citet{2003astro.ph..1539K}, \citet{2015ASPC..493...65K}, and \citet{2019A&ARv..27....7C}.

The second noteworthy phenomenon is associated with cool ZZ~Ceti stars: the presence of so-called outbursts in their light curves \citep{2015ApJ...809...14B}. During an outburst, the mean brightness of the star increases by a few percent on an approximately hourly timescale. The star then remains in this brightened state for several hours before fading back to its original brightness. The recurrence of outbursts does not follow a regular pattern: they may recur after days or weeks, but longer intervals are also possible \citep{2023A&A...674A.204B}. Importantly for the present work, the outbursting DAVs KIC~4552982 and PG~1149+057 are also rotation detections in the asteroseismic sense: both have stable pulsation splittings interpreted as rotational, with reported periods of $17.47\pm0.04$~h and about 1.2~d, respectively \citep{2015ApJ...809...14B,2015ApJ...810L...5H}. From a theoretical point of view, outbursts may be caused by non-linear mode coupling, radiative damping, or phase shifts of travelling waves reflected from the outer turning point close to the convection zone; see, for example, \citet{2017ASPC..509..303B}, \citet{2018ApJ...863...82L}, and \citet{2020ApJ...890...11M}. Most recently, \citet{2026arXiv261028513G} showed that resonant three-mode coupling can place DAV pulsations in periodic or chaotic regimes, suggesting that deterministic chaos and intermittency may contribute to the irregular amplitudes and recurrence times of outbursts. We return to the outburst candidates found in individual light curves later in the paper.

The present study is based on observations obtained by the Transiting Exoplanet Survey Satellite (TESS) \citep{2015JATIS...1a4003R}. Launched on 18 April 2018, TESS made it possible to observe selected short-period pulsators with a 120-second cadence. The 2020–2022 Extended Mission provided an even more valuable observing mode: 20-second cadence data became available, proving particularly suitable for the study of pulsating white dwarfs. The second Extended Mission (EM2), beginning with Sector~56 in 2022, continued the short-cadence and ultrashort-cadence monitoring that is essential for resolving ZZ~Ceti pulsations in TESS data. Studies of pulsating white dwarfs and subdwarfs observed by TESS are coordinated by the TESS Asteroseismic Science Consortium Compact Pulsators Working Group (TASC WG\#8)\footnote{\url{https://tasoc.dk/wg8/}}.

The first paper in this series, published in 2024, focused on the rotation of ZZ~Ceti stars using TESS data. The present paper is a continuation of that work, incorporating data from newly observed sectors. In addition to searching for multiplets indicative of stellar rotation, we also search for possible outburst candidates in order to check for any evidence of a connection between the outburst events and the rotation of the host star. Because two outbursting DAVs with splitting-based rotation estimates are already known, our aim is to determine whether the present homogeneous TESS analysis adds further examples or yields a measurable trend.

\section{The TESS data sets}
\label{sect:analyses}

We constructed the target list from three published ZZ~Ceti samples: Table~4 of \citet{2016IBVS.6184....1B}, the 75 DAVs analysed by \citet{2022MNRAS.511.1574R}, and the 32 new TESS ZZ~Ceti stars reported by \citet{2025ApJ...984..112R}, hereafter R25. The combined list contains 277 stars, of which 239 have available TESS light curves up to Sector~96. All 239 targets have 120~s cadence data, and 178 also have 20~s cadence observations.

We used the Science Processing Operations Centre (SPOC) light curve Flexible Image Transport System (FITS) products downloaded from the Mikulski Archive for Space Telescopes (MAST) and adopted the systematics-corrected Pre-search Data Conditioning Simple Aperture Photometry flux (\texttt{PDCSAP\_FLUX}) values as the starting point of the analysis. After removing invalid files, the data set contained 2201 sector light curves, consisting of 1343 120~s and 858 20~s cadence files.

For each sector light curve, we extracted the time, flux, flux uncertainty, and quality flag, removed invalid rows, and corrected long-term trends with cadence-dependent smoothing windows. The corrected sector light curves were then merged separately for each target and cadence, yielding 417 combined light curves: 239 at 120~s cadence and 178 at 20~s cadence.

We performed a Fourier analysis and iterative pre-whitening on all combined light curves using Lomb-Scargle periodograms computed with \textsc{Astropy} \citep{2013A&A...558A..33A,2018AJ....156..123A,2022ApJ...935..167A} and sinusoidal fitting. The iteration was stopped when no peak exceeded the adopted 0.1 percent false-alarm probability threshold or when 99 frequencies had been extracted. The search was carried out up to 360\,d$^{-1}$ for the 120~s data and 900\,d$^{-1}$ for the 20~s data.

Quality control was carried out at the sector and point levels. Sectors with unusually large scatter relative to other sectors of the same target and cadence were removed, affecting 18 sectors in 14 combined light curves. We then applied an iterative outlier rejection at $8$ median absolute deviations (MADs) to the merged light curves. Here, MAD is defined as $\operatorname{median}(|x_i-\operatorname{median}(x)|)$; it is a robust measure of scatter because it is much less sensitive to outliers than the standard deviation. Only the modified light curves were reanalyzed. The final frequency lists, residual light curves, and diagnostic periodograms form the basis of the triplet search described below. The complete processing sequence was therefore: target-list construction, FITS light curve extraction, cadence-separated correction and merging, sector- and point-level quality control, frequency analysis with iterative prewhitening, automated triplet search, and final visual validation.

\section{Searching for triplets}
\label{sect:multiplets}

We searched for rotationally split triplets using the frequency lists obtained from the final pre-whitening analysis. In total, 417 frequency lists were analysed, corresponding to 239 light curves with 120~s cadence and 178 light curves with 20~s cadence. The search was automated to make the first candidate selection reproducible, but the final acceptance of triplets was still based on visual inspection.

The search was restricted to frequencies above 25\,d$^{-1}$, and only the 20 highest-amplitude frequencies from each list were used to reduce the number of chance combinations. The algorithm then searched for nearly symmetric frequency triples
\[
f_1,\ f_2,\ f_3
\]
satisfying
\[
f_2 - f_1 \simeq f_3 - f_2 .
\]
The mean frequency separation of a candidate triplet is defined as
\[
\Delta f = \frac{1}{2}\left[(f_2-f_1) + (f_3-f_2)\right].
\]

Candidate triplets with similar splittings were grouped into consensus clusters and ranked. A consensus cluster is a set of automatically detected triplet candidates whose mean frequency separations agree within the tolerance set by the frequency resolution of the corresponding light curve. Thus, a cluster represents a common splitting value supported by one or more independent triplet patterns, rather than a single isolated frequency triplet. The ranking favoured clusters with more candidate members, more peak-disjoint members, a larger number of unique peaks, and smaller scatter in the inferred splittings. This step helped to distinguish cases where several modes suggested the same rotational splitting from isolated, accidental frequency triples. We also generated doublet candidates as a diagnostic product, but only complete triplets were used for the rotation periods presented here.

The automated search identified triplet candidates in 44 frequency lists of 30 TIC targets. These candidates were inspected visually using frequency-amplitude diagrams and zoomed periodograms. Closely spaced alternative peaks were treated as competing detections of the same triplet component, and the highest-amplitude peak was accepted. The triplet search can be summarised numerically as 417 analysed frequency lists, 44 frequency lists with automatically identified triplet candidates in 30 candidate TIC targets, and 11 accepted rotators after visual inspection. 

A natural interpretation of the detected triplet patterns is that they correspond to rotationally split components of $\ell=1$ non-radial pulsation modes. Once the frequency separation between the components, $\delta\nu$, is measured, it can be used to estimate the stellar rotation period.

In the limit of slow rotation, the first-order frequency shift of a non-radial mode is given by
\begin{equation}
\label{eq:rot}
\delta \nu_{k,\ell,m} = m (1-C_{k,\ell}) \Omega ,
\end{equation}
\noindent
where $k$, $\ell$, and $m$ denote the radial order, spherical degree, and azimuthal order of the pulsation mode, respectively. The quantity $C_{k,\ell}$ is the Ledoux constant, which approaches $C_{k,\ell} \approx 1/[\ell(\ell+1)]$ for high-overtone ($k\gg\ell$) $g$-modes, and $\Omega$ is the uniform rotation frequency.

As noted, for example, by \citet{2017ApJS..232...23H}, the observed doublet and triplet components in DAV stars cannot always be assumed to originate from high-overtone modes. To allow a direct comparison with the rotation rates reported by \citet{2017ApJS..232...23H}, we therefore adopted a uniform value of $C_{k,\ell}=0.47$ when estimating the rotation periods of all stars in our sample.

We do not quote formal uncertainties for the derived rotation periods. If the Ledoux constant is kept fixed, the fractional uncertainty of the rotation period follows directly from the uncertainty of the measured splitting. In practice, however, the dominant uncertainties are systematic rather than statistical, arising from the TESS frequency resolution, amplitude and frequency variability, the identification of the triplet components, and the assumed value of $C_{k,\ell}$. As a representative scale, the Rayleigh frequency resolution $1/\Delta T$ of the accepted triplet light curves ranges from about 0.005 to 0.21\,$\mu$Hz, with a median of about 0.011\,$\mu$Hz for the entries in Table~\ref{tab:triplet_tess_data}. For the five stars in which the same triplet was measured independently in the 120~s and 20~s data, the adopted mean splittings from the two cadences agree within 0.016\,$\mu$Hz. The periods in Table~\ref{tab:triplet_parameters} should therefore be interpreted as estimates based on the measured mean splittings.

Tables~\ref{tab:triplet_tess_data} and \ref{tab:triplet_parameters} summarise the accepted triplet sample. Table~\ref{tab:triplet_tess_data} lists the TESS observations used for the accepted rotators, while Table~\ref{tab:triplet_parameters} combines the selected central triplet frequencies, adopted mean splittings, estimated rotation periods, stellar masses, and literature comparisons. The 11 independent rotation estimates correspond to 11 different stars. Objects previously reported in B24 or R25 are marked accordingly.

\begin{table*}[t]
\center
\setlength{\tabcolsep}{3pt}
\caption{TESS data used for the accepted triplet candidates.}
\label{tab:triplet_tess_data}
\begin{tabular}{rrrr r p{6.0cm} c}
\hline
\hline
\multicolumn{1}{c}{TIC ID} & Start time & $N$ & $\Delta T$ & $G$ mag & Sector & CROWDSAP \\
       & \multicolumn{1}{c}{[BTJD]} & & [d] & & [US: 20 s, S: 120 s] & \\
\hline
 11626063 & 1683.358 & 60777   & 1849.822 & 15.149 & S14, S41, S54, S81 & 0.075 \\
 55650407 & 2036.278 & 2582841 & 1896.846 & 14.991 & US27--39, US61--69, US87--89, US93--96 & 0.968 \\
101014997 & 1385.938 & 59708   & 784.008  & 15.064 & S3--4, S30--31 & 0.921 \\
101014997 & 2115.887 & 191458  & 54.060   & 15.064 & US30--31 & 0.919 \\
149863849 & 2361.777 & 49616   & 1493.320 & 13.556 & S39, S66, S93 & 0.146 \\
149863849 & 2361.776 & 289393  & 1493.318 & 13.556 & US39, US66, US93 & 0.146 \\
199666369 & 2390.654 & 2585915 & 1272.173 & 12.291 & US40--41, US47--60, US73--86 & 0.976 \\
217329125 & 2825.262 & 207304  & 759.118  & 15.777 & US56, US83 & 0.935 \\
262872628 & 2333.857 & 509050  & 1521.147 & 13.443 & US38--39, US65--66, US93 & 0.710 \\
287926830 & 2825.261 & 308895  & 759.118  & 15.962 & US56, US82--83 & 0.191 \\
313109945 & 1683.352 & 167583  & 1085.625 & 18.591 & S14--15, S20--22, S40--41, S47--48, S53 & 0.077 \\
313109945 & 2390.651 & 506832  & 378.326  & 18.591 & US40--41, US47--48, US53 & 0.077 \\
343296348 & 1629.139 & 75537   & 2225.498 & 15.869 & S12--13, S39, S66, S93 & 0.394 \\
343296348 & 2361.774 & 265161  & 1492.863 & 15.869 & US39, US66, US93 & 0.394 \\
394015496 & 1325.300 & 73551   & 2555.689 & 15.811 & S1, S27--28, S67, S94 & 0.256 \\
394015496 & 2061.855 & 246532  & 1819.135 & 15.811 & US28, US67, US94 & 0.256 \\
\hline
\end{tabular}
\begin{flushleft}\footnotesize
Notes: Start times are given as Barycentric TESS Julian Date (BTJD) = Barycentric Julian Date (BJD)$-2457000$. $N$ is the number of data points after cleaning the light curve, $\Delta T$ is the total length of the data sets including gaps, and the sector labels indicate the cadence: S denotes 120~s cadence data, while US denotes 20~s ultrashort-cadence data. CROWDSAP is a crowding metric, the ratio of the target flux to the total flux in the aperture. It is a SPOC pipeline estimate, based on the TESS Input Catalog and the adopted photometric aperture.
\end{flushleft}
\end{table*}

\begin{table*}[!t]
\centering
\caption{Accepted rotational-triplet parameters.}
\label{tab:triplet_parameters}
\begingroup
\begin{tabular}{rrrrrrlr}
\hline
\hline
TIC ID & Cad. & $f_0$ & $\langle\Delta f\rangle$ & $P_{\rm rot}$ & Mass & Status & Lit. $P_{\rm rot}$\\
 & & [$\mu$Hz] & [$\mu$Hz] & [h] & [$M_\odot$] & & [h]\\
\hline
 11626063 & 120 s & 3902.746 & 1.560  & 94.35 & 0.597 & new & \\
 55650407 &  20 s & 4997.882 & 14.227 & 10.35 & 0.574 & R25 & 12.12\\
101014997 & 120 s & 1618.399 & 12.414 & 11.86 & 0.623 & B24 & 11.9\\
101014997 &  20 s & 1618.372 &        &       &       &     & \\
149863849 & 120 s & 1496.014 & 18.207 & 8.09  & 0.657 & R25 & 14.49--22.73\\
149863849 &  20 s & 1496.003 &        &       &       &     & \\
199666369 &  20 s & 9150.865 & 16.139 & 9.12  & 0.783 & B24 & 9.1\\
217329125 &  20 s & 3897.760 & 5.466  & 26.93 & 0.614 & new & \\
262872628 &  20 s & 5191.850 & 12.985 & 11.34 & 0.674 & B24 & 12.2\\
287926830 &  20 s & 3618.345 & 5.537  & 26.59 & 0.572 & new & \\
313109945 & 120 s & 3347.356 & 1.692  & 87.03 & 0.410 & B24 & 85.5\\
313109945 &  20 s & 3347.356 &        &       &       &     & \\
343296348 & 120 s & 3475.054 & 6.108  & 24.10 & 0.586 & R25 & 12.35\\
343296348 &  20 s & 3475.070 &        &       &       &     & \\
394015496 & 120 s & 3228.026 & 5.030  & 29.27 & 0.607 & B24 & 29.4\\
394015496 &  20 s & 3228.027 &        &       &       &     & \\
\hline
\end{tabular}
\endgroup
\begin{flushleft}\footnotesize
Notes: From left to right, the columns give the TESS Input Catalog identifier (TIC ID); the light curve cadence (Cad.; 120 or 20~s); the frequency of the middle triplet component ($f_0$) measured at that cadence; the adopted mean splitting for the star ($\langle\Delta f\rangle$; averaged between cadences when both are available); the rotation period ($P_{\rm rot}$) derived from this splitting assuming $C_{k,\ell}=0.47$; the white dwarf mass from the Gaia EDR3 hydrogen-atmosphere photometric solution (\citealt{2021MNRAS.508.3877G}); the study status (new -- reported in this work the first time; B24 -- reported by \citealt{2024A&A...684A..76B}; R25 -- reported in R25); and the previously published rotation period, if available. For stars measured at both cadences, the star-related quantities are shown only in the first row of the TIC entry.
\end{flushleft}
\end{table*}

\begin{figure*}[ht]
\centering
\includegraphics[width=0.81\textwidth]{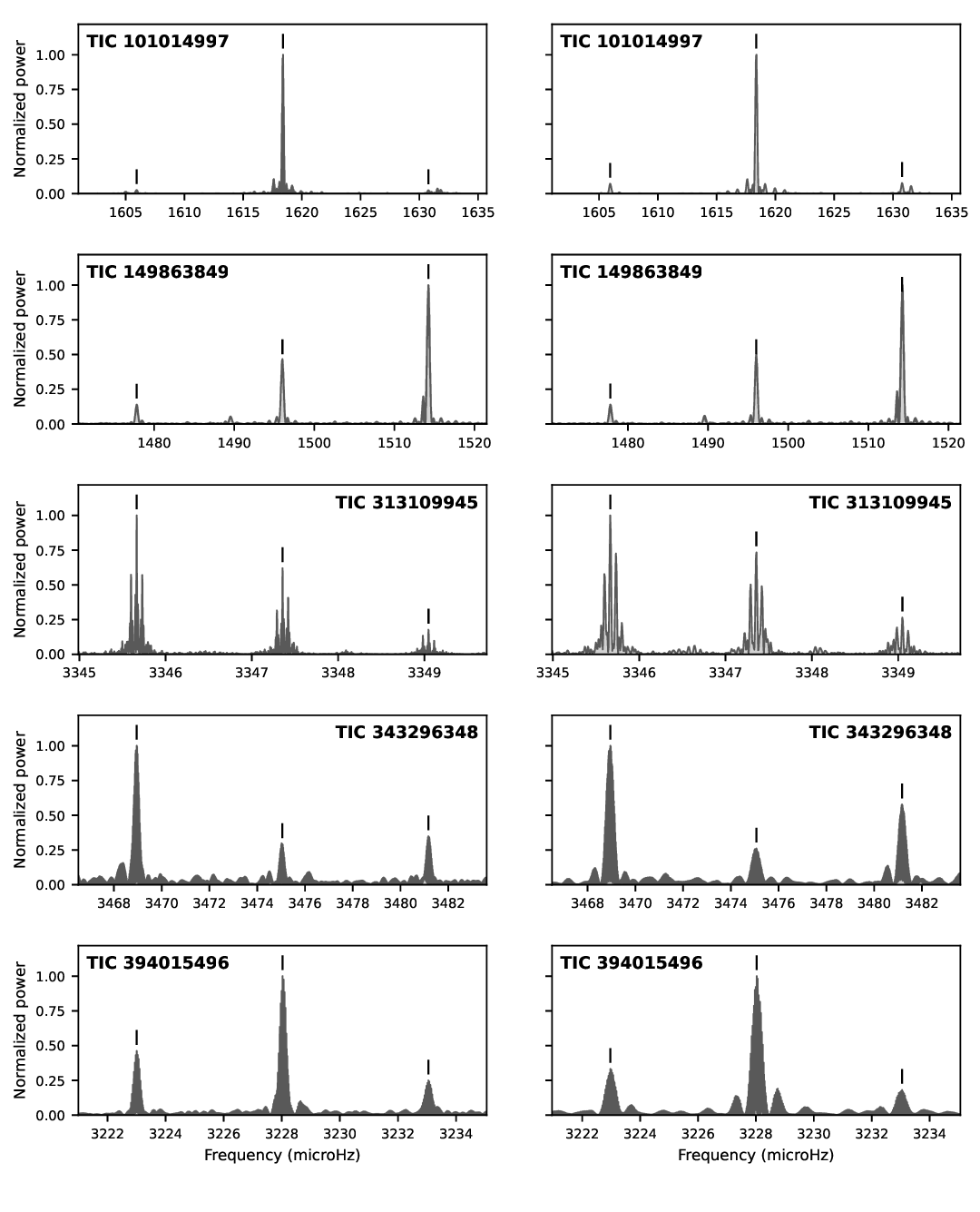}
\caption{Periodograms of the accepted rotationally split triplet candidates observed in both 120~s and 20~s cadence data. For each star, the left and right panels correspond to the 120~s and 20~s cadence data, respectively. The horizontal frequency scale is given in $\mu$Hz, and the periodogram power is normalised independently in each panel. Short vertical black markers above the peaks indicate the selected triplet components.}
\label{fig:triplet1}
\end{figure*}

\begin{figure*}[t]
\centering
\includegraphics[width=0.81\textwidth]{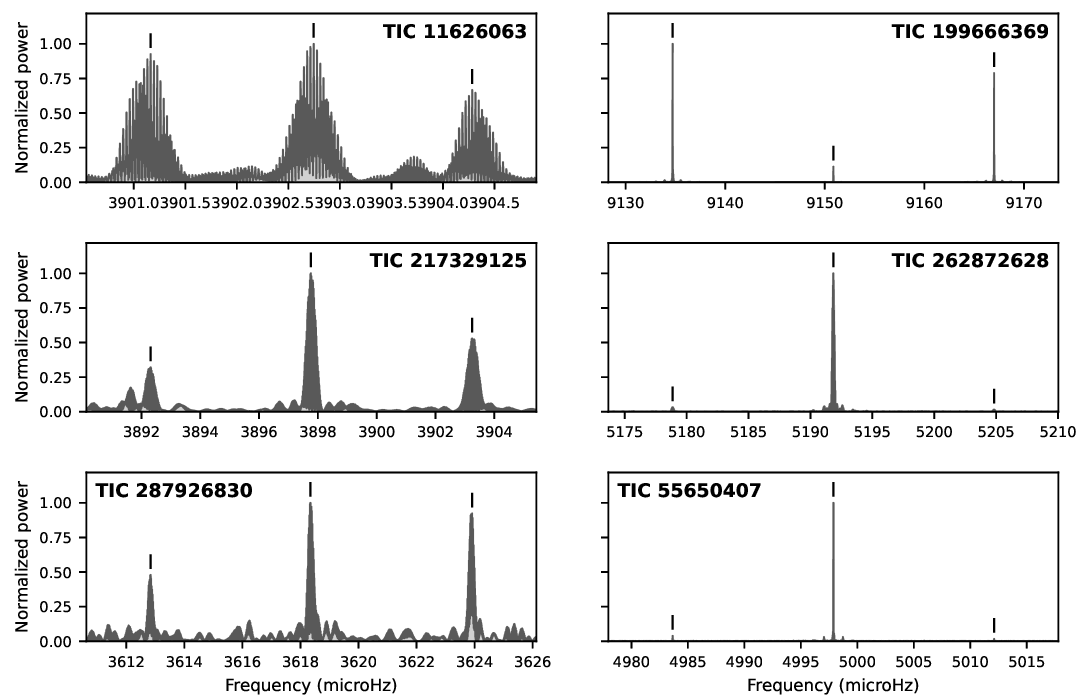}
\caption{Same as Fig.~\ref{fig:triplet1} but for the remaining accepted triplet candidates with data available at only one cadence. The cadence for each star is listed in Table~\ref{tab:triplet_tess_data}.}
\label{fig:triplet2}
\end{figure*}

The final visually accepted triplets are shown in Figs.~\ref{fig:triplet1} and \ref{fig:triplet2}. These panels show the local periodogram structure around the selected components and provide visual confirmation for the component identifications used in Table~\ref{tab:triplet_parameters}.

\section{Investigation of outburst candidates}

We searched for outburst-like brightenings in the corrected TESS light curves, treating the 120 s and 20 s cadence data sets separately. Before the search, we removed problematic sectors identified from the sector-quality diagnostics and applied an iterative outlier rejection using an \(8\,{\rm MAD}\) threshold in order to suppress isolated deviant points without removing genuine multi-hour brightenings. The outburst search was deliberately conservative; it was designed to identify a robust candidate sample rather than to measure a complete occurrence rate.

The automatic outburst search was based on the time-domain behaviour of the corrected fluxes. We first identified contiguous groups of points lying more than \(3\,{\rm MAD}\) above the median level of the light curve. To reject low-level fluctuations and isolated excursions, a group was kept only if its maximum point exceeded \(5\,{\rm MAD}\), if it contained enough consecutive high points for the corresponding cadence, and if its duration was shorter than 0.30\,d. This upper duration limit is longer than the individual events ultimately measured here, but it excludes broad residual trends that are unlikely to be classical outburst profiles. The automatic candidates were then ranked using a score that combines the peak significance, the number of high points, and the integrated excess above the local quiescent flux level.

For the final duration measurements, we used a separate local procedure. Around each candidate event, we estimated the local quiescent flux level from the median flux outside the event and estimated the local scatter as \(1.4826\,{\rm MAD}\), which is the Gaussian-equivalent robust standard deviation. The light curve was smoothed with a 0.25\,h running mean before measuring the event width. This smoothing scale is much longer than the 20 s and 120 s cadences, so it suppresses point-to-point noise and short pulsation-cycle fluctuations, but it is still shorter than the hour-scale brightenings that we aimed to measure. The duration was defined as the interval over which the smoothed light curve remained above the threshold given by the local quiescent flux level plus $3\sigma$. This definition gives a reproducible, conservative width for each event; it should be regarded as the duration of the high-confidence brightened part of the light curve rather than necessarily the full physical outburst duration.

The automatic search produced 39 event candidates in 13 targets. After visual inspection of the local light curves, sector context, and smoothed profiles, we accepted 19 outburst-like events in 11 stars for further validation; their pre-validation light curve profiles are provided in the electronic material. Thus, about half of the automatic triggers were rejected at this validation stage, mostly because they were too close to residual trends, sector-boundary features, or ambiguous low-amplitude variability.
The visual rejection fraction provides empirical evidence that automatic thresholding alone is insufficient and that even the light-curve-selected sample requires additional validation before being interpreted astrophysically.

Because moving Solar System objects can produce short, outburst-like brightenings in TESS apertures, as demonstrated for HZ~4 by \citet{2023RNAAS...7..156B}, we inspected the target-pixel files around the selected events using \textsc{Lightkurve} \citep{2019AAS...23310908B}. For each event, we compared the PDCSAP flux, the Simple Aperture Photometry (SAP) background level, and event-centred difference images. These diagnostics show compact moving residuals or suspicious background/aperture behaviour for TIC 0001116746, 101901770, 1103347368, 366681966, 67441210, 800153845, 800562727, and 842051075. We show a representative example of the diagnostics indicative of asteroid contamination or otherwise considered suspicious in Appendix~\ref{app:asteroid_pixel_diagnostics}; the full diagnostic set is included in the electronic material.

We then queried the Minor Planet Center MPChecker service at the mid-time of each suspicious event, using the TESS observatory code C57, a 5 arcmin search radius, and a limiting magnitude of $V=24$. MPChecker returned at least one known minor planet within the search radius for all 11 event/cadence queries associated with the eight rejected targets. The nearest matches are provided as electronic material at the Centre de Donn\'ees astronomiques de Strasbourg (CDS) via the VizieR service (see Table~\ref{tab:mpchecker_matches}). These matches support the pixel-level interpretation that the affected brightenings are more likely due to moving-object contamination than to stellar outbursts. We therefore reject these eight events as stellar outburst candidates. The three objects without clear moving-residual signatures in the inspected frames are TIC 219442838, TIC 307908863, and TIC 630426257 (Appendix~\ref{app:outburst_pixel_diagnostics}). The journal of observations for the 11 stars that passed the light curve validation stage is given in Table~\ref{tab:outburst_observations_detected}.

\begin{table*}[!t]
\centering
\setlength{\tabcolsep}{3pt}
\caption{Nearest MPChecker matches for the rejected events showing signatures of asteroid contamination or other suspicious features.}
\label{tab:mpchecker_matches}
\begin{tabular}{rrrlcc}
\hline\hline
TIC ID & Sector & Query JD & Nearest MPC object & $V$ [mag] & Offset [arcmin] \\
\hline
0001116746 & 48 & 2459620.368768 & (49513) 1999 CK28 & 17.6 & 0.2E, 0.1S \\
\ldots \\
\hline
\end{tabular}
\begin{flushleft}\footnotesize
Notes: The full table is available electronically only at the CDS via the VizieR service. The query time is the midpoint of the brightening event in Julian Date. Searches used observatory code C57, a 5 arcmin radius, and limiting magnitude $V=24$. Offsets are MPChecker offsets in arcminutes relative to the target position. Multiple rows for the same TIC correspond to different cadence or sector event/cadence queries.
\end{flushleft}
\end{table*}

\begin{table*}[!t]
\centering
\caption{Log of the TESS observations in which outburst-like brightenings were detected.}
\label{tab:outburst_observations_detected}
\begin{tabular}{rrcccccc}
\hline\hline
TIC ID & Cadence [s] & Sector & Start time & $N$ & $\Delta T$ [d] & $G$ [mag] & CROWDSAP \\
\hline
  1116746 & 120 & 48 & 2609.857 & 14931 & 26.13 & 16.467 & 0.803 \\
 67441210 & 120 & 20 & 1842.509 & 16016 & 26.32 & 15.823 & 0.672 \\
101901770 & 120 & 64 & 3041.289 & 16698 & 24.75 & 16.093 & 0.771 \\
219442838 &  20 & 63 & 3015.228 & 99445 & 24.51 & 16.415 & 0.189 \\
          &  20 & 90 & 3747.164 & 97018 & 25.73 & 16.415 & 0.178 \\
          & 120 &  9 & 1544.323 & 15294 & 24.16 & 16.415 & 0.186 \\
          & 120 & 63 & 3015.228 & 16552 & 24.46 & 16.415 & 0.189 \\
          & 120 & 90 & 3747.166 & 16089 & 25.72 & 16.415 & 0.178 \\
307908863 &  20 & 48 & 2610.418 & 87422 & 25.57 & 15.559 & 0.138 \\
366681966 & 120 & 72 & 3262.249 & 14274 & 23.32 & 14.980 & 0.983 \\
630426257 & 120 & 70 & 3208.600 & 14432 & 24.39 & 16.955 & 0.435 \\
800153845 &  20 & 44 & 2500.363 & 89161 & 24.08 & 17.322 & 0.123 \\
          & 120 & 44 & 2500.366 & 14910 & 24.07 & 17.322 & 0.123 \\
          & 120 & 61 & 2963.923 & 14806 & 23.83 & 17.322 & 0.145 \\
800562727 & 120 & 44 & 2500.367 & 14737 & 24.07 & 17.702 & 0.343 \\
842051075 & 120 & 62 & 2989.520 & 15551 & 24.64 & 18.232 & 0.191 \\
1103347368&  20 & 51 & 2698.355 & 39187 & 19.19 & 15.829 & 0.124 \\
          & 120 & 51 & 2698.360 &  6499 & 19.18 & 15.829 & 0.124 \\
\hline
\end{tabular}
\begin{flushleft}\footnotesize
Notes: Start times are given as BTJD = BJD $-2457000$. The number of points and the time base refer to the corrected sector light curves used in the analysis. For targets with multiple entries, the TIC ID is listed only in the first row. Pixel-level diagnostics show that only a subset of these events is accepted as candidate stellar outbursts.
\end{flushleft}
\end{table*}

\begin{figure*}[!t]
\centering
\includegraphics[width=0.95\textwidth]{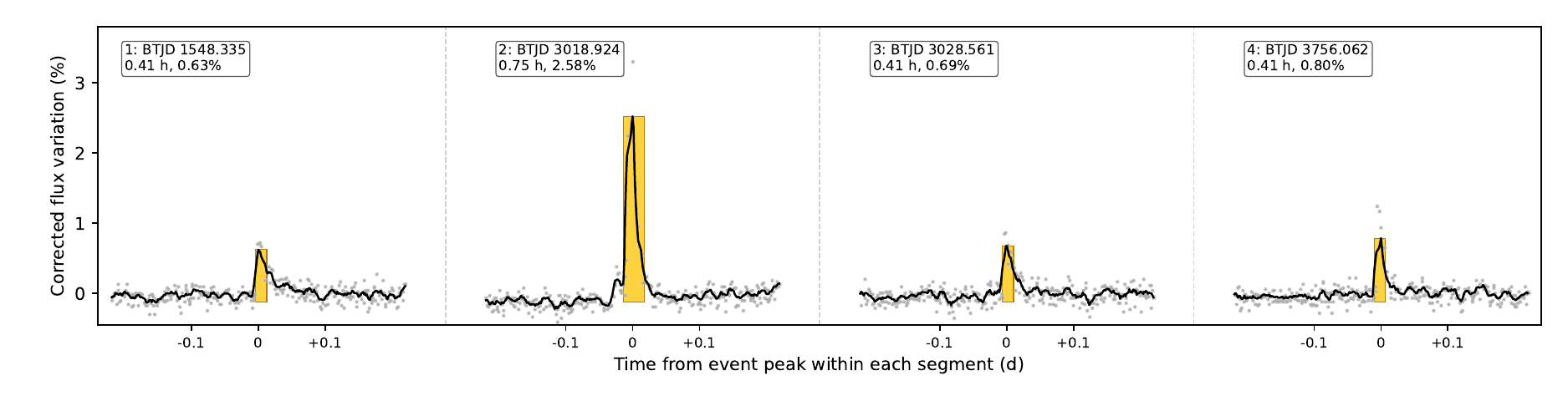}
\caption{Overview of the four accepted 120~s cadence outburst-like brightenings detected in TIC 219442838. Each segment is centred on the peak of one event and is plotted on the same relative time scale. Grey points show the corrected TESS fluxes, while the black curve shows a 7-point running mean. Yellow shading marks the measured brightened interval. The labels give the BTJD peak time, measured duration, and peak flux excess of each event. The recurrence of similar short brightenings in different TESS sectors supports the interpretation of TIC 219442838 as a candidate stellar outburster but does not by itself prove that the events are classical ZZ~Ceti outbursts.}
\label{fig:tic219_outburst_overview}
\end{figure*}

\begin{table*}[!t]
\centering
\caption{Accepted outburst-like brightening events identified in the TESS data.}
\label{tab:outburst_candidates}
\begin{tabular}{rccrccc}
\hline\hline
TIC ID & $T_{\mathrm{eff}}$ & $\log g$ & Cadence & Peak time & Duration & Peak flux \\
 & [K] & [cgs] & [s] & [BTJD] & [hr] & [\%] \\
\hline
219442838 & 11609 & 8.01 & 120 & 1548.335 & 0.41 & 0.63 \\
 &  &  & 20 & 3018.921 & 0.47 & 3.90 \\
 &  &  & 120 & 3018.924 & 0.75 & 2.58 \\
 &  &  & 120 & 3028.561 & 0.41 & 0.69 \\
 &  &  & 20 & 3756.060 & 0.23 & 1.14 \\
 &  &  & 120 & 3756.062 & 0.41 & 0.80 \\
307908863 & 12042 & 8.03 & 20 & 2612.149 & 0.33 & 0.88 \\
630426257 & 12258 & 7.85 & 120 & 3224.991 & 0.84 & 2.23 \\
\hline
\end{tabular}
\begin{flushleft}\footnotesize
Notes: Effective temperatures and surface gravities are taken from the Gaia EDR3 white dwarf catalogue. For TIC~219442838, multiple events are listed. Peak times are given as BTJD = BJD $-2457000$, durations were measured from the smoothed light curves, and peak fluxes are the largest smoothed excesses relative to the local quiescent flux level.
\end{flushleft}
\end{table*}

\section{Discussion}
\label{sect:discussion}

In the following section, we first discuss our findings on the rotational frequency-splittings, then the identified outburst events; finally, we investigate the relation between the two phenomena, also including literature examples.

\subsection{Rotational multiplets}
\label{sect:discussion-rotmultiplets}

\begin{figure}[htbp]
%\centering
\includegraphics[width=0.45\textwidth]{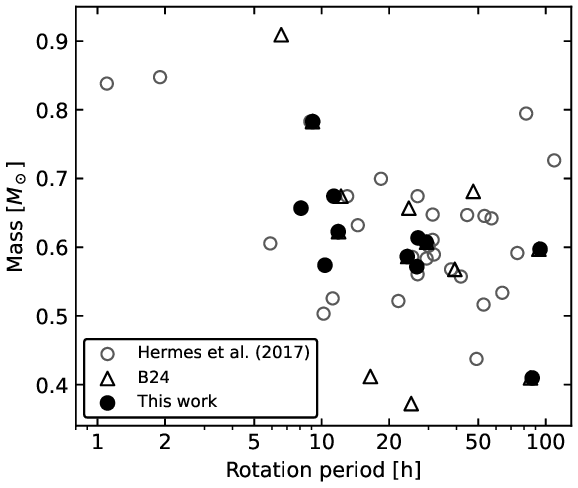}
\caption{Stellar mass as a function of rotation period for the ZZ~Ceti stars with accepted triplet frequency structures. Filled black circles show the objects analysed in this work, open triangles show the B24 rotators, and open circles show the comparison objects from \citet{2017ApJS..232...23H} for which Gaia EDR3 white dwarf catalogue masses are available. All masses are the hydrogen-atmosphere photometric solutions from the Gaia EDR3 white dwarf catalogue. The horizontal axis is logarithmic.}
\label{fig:mass}
\end{figure}

Similarly to B24, we plotted the stellar mass as a function of the rotation period for the accepted triplet candidates in Fig.~{\ref{fig:mass}}. Our new results are shown as filled black circles. We also included the 13 rotators from B24 for which Gaia EDR3 white dwarf catalogue masses were available; these are shown as open triangles. For comparison, we plotted the isolated pulsating white dwarfs from \citet{2017ApJS..232...23H} (their Table 4) as open circles, excluding EPIC~201730811, similarly to their original figure, because it is a close post-common-envelope binary. The horizontal axis shows the rotation period in hours on a logarithmic scale, and the vertical axis shows the stellar mass in solar units. Thus, all masses shown in the figure are taken consistently from the Gaia EDR3 white dwarf catalogue.

The comparison with the \citet{2017ApJS..232...23H} sample illustrates the previously noted tendency that the most massive white dwarfs appear to rotate faster. Our newly identified rotators do not include very high-mass objects, and therefore they do not strongly test this trend at the high-mass end. Within our lower-mass sample, the points show substantial scatter and may even suggest longer rotation periods toward higher masses. However, given the small number of objects, the limited mass range, and the uncertainties inherent in deriving rotation periods from a small number of multiplets, we do not interpret this as evidence for an opposite correlation. Rather, our results are broadly consistent with the large dispersion seen among normal-mass white dwarfs in the \citet{2017ApJS..232...23H} sample.

A complementary comparison is provided by the photometric white dwarf rotation census of \citet{2024ApJ...974..314O}. They analysed TESS light curves of white dwarfs up to Sector~69, used \texttt{TESS\_LOCALIZE} to test the origin of the signals, and found a median period of 3.9~h for their high-probability likely single white dwarfs. This distribution is much shorter than the 24.2~h median of their compiled asteroseismic sample. This difference should not be interpreted as a direct conflict with our triplet-based periods. The photometric periods trace stable surface inhomogeneities, probably often connected with magnetic or chemical surface structure, whereas rotational splittings measure the rotation sampled by non-radial pulsation modes and are usually interpreted as an internal or mode-weighted rotation rate. Our accepted triplet periods of approximately 8--94~h therefore remain most directly comparable to the seismological DAV sample, not to the full photometric surface-rotation census. For the same reason, the absence of overlap between our outburst candidates and triplet rotators cannot be converted into a test of any possible relation between outbursts and photometric surface rotation. Such a test would require surface-rotation detections and pixel-level validated outburst classifications for the same stars.

Measurements of both kinds do exist for a few individual pulsating white dwarfs. For the DBV PG~0112+104, \citet{2017ApJ...835..277H} measured a 10.174~h photometric modulation attributed to a surface inhomogeneity and rotationally split pulsation modes broadly consistent with a similar rotation timescale. In contrast, \citet{2024ApJ...974..314O} found low-frequency photometric variability in the DAV G226-29 (TIC~199666369), one of our accepted triplet rotators, whose 9.12~h splitting-based period differs from their variable signal at 0.93--3.25~h. Although a surface-rotation interpretation, including differential rotation, is possible, the changing low-frequency signal has not been reliably identified as a surface-rotation period. These individual examples do not establish a general relation between surface and seismological rotation, and the very different period distributions of the photometric and seismological samples cannot by themselves be interpreted as radial differential rotation because the samples have different selection criteria.

The automated search described above was designed to make the identification of rotationally split triplets reproducible across a heterogeneous TESS data set. Nevertheless, the number of accepted triplet candidates is modest: we accepted 11 stars out of 239 TESS-observed ZZ~Ceti targets, corresponding to an apparent detection efficiency of only 4.6 percent for the strict identification of complete triplets. If all objects with automated triplet candidates are counted before visual validation, the fraction is 30 out of 239 targets, or 12.6 percent. These numbers should not be interpreted as the intrinsic occurrence rate of rotational splitting among DAVs. They instead describe the efficiency with which complete, sufficiently clean triplet patterns can be recovered from the available TESS light curves using our conservative criteria. Several observational and methodological effects reduce this efficiency.

We note that our automated search was not designed to recover every previously reported rotational splitting. In B24, as well as in the samples of \citet{2022MNRAS.511.1574R} and R25, some rotation estimates were based on doublets or on multiplet components identified through detailed visual inspection of individual stars. In the present work, we required complete triplet candidates in the automatically generated frequency lists, and the pre-whitening was stopped when no peak exceeded the adopted 0.1 percent false-alarm probability threshold. Consequently, cases in which one component of a previously reported triplet falls below this threshold, is blended with nearby variability, or is absent in the currently analysed TESS sectors may not be recovered by our pipeline. 

The observed distinction between triplets, doublets, singlets, and non-detections is therefore sensitive to the target-specific noise level and window function. A physically present triplet could appear as a doublet, a singlet, or no accepted multiplet in noisier data if one or more components fall below the adopted significance threshold. Degrading high signal-to-noise Kepler/K2 light curves to TESS-like noise levels would be a useful dedicated recovery experiment, which would require a separate homogeneous simulation of mode amplitudes, window functions, and target-dependent TESS noise. Such a simulation is beyond the scope of the present observational paper. The non-recovery of some literature splittings should therefore not be interpreted as evidence against those detections, but rather as a consequence of the stricter and more homogeneous selection criteria adopted here.

First, rotational splitting can only be recognised when at least three components of the same mode are detected with sufficient signal-to-noise ratio and frequency resolution. Many ZZ~Ceti stars show only a small number of significant peaks in the TESS bandpass, while in others, one or two components of a multiplet may fall below the detection threshold because of geometric cancellation, amplitude variability, or the inclination of the pulsation axis. In such cases, the star may still be rotating at a measurable rate, but the available data do not contain a complete triplet. This is why we also generated doublet candidates as a diagnostic product, although we did not use doublets alone for the rotation-period determinations presented here.

Second, the time coverage is highly non-uniform. Some targets were observed in many sectors, while others have only one or a few sectors of data. Shorter time bases reduce the Rayleigh frequency resolution and mainly affect cases in which the rotational splitting is comparable to or only a few times larger than $1/\Delta T$. For the accepted triplets in this paper, the measured splittings are 1.56--18.21\,$\mu$Hz, whereas the Rayleigh resolutions of the corresponding light curves are at most 0.21\,$\mu$Hz and are usually much smaller. Thus, the accepted triplets are not limited by frequency resolution in the simple sense of being unresolved; the resolution issue is instead relevant to rejecting marginal, blended, or slowly rotating candidates that did not enter the final accepted sample. Conversely, very dense frequency spectra can also be problematic: cool DAVs often show amplitude and frequency variability, unresolved close peaks, and possible combination frequencies, all of which can generate many formally acceptable frequency triples. The ranking and consensus-clustering steps reduce this problem but cannot eliminate it completely.

The main false-positive mechanisms for triplet candidates are, therefore, accidental near-equidistant frequency combinations, unresolved amplitude or frequency modulation, aliases produced by the observing window, and combination frequencies that mimic symmetric structures. Additional caution is required for frequency lists that have reached the maximum pre-whitening limit because these objects may have complex residual spectra in which chance alignments become more likely. For this reason, the automated stage was used only to generate and rank candidates. The final sample was restricted to cases where the proposed triplet structure remained plausible after inspection of the frequency-amplitude diagrams, zoomed periodograms, and the consistency of the inferred splittings.

\subsection{Outbursts}

The outburst search is subject to a different set of biases. TESS provides long, homogeneous light curves, but the large pixel size, flux dilution, sector gaps, and target-to-target differences in contamination can strongly affect the detectability of low-amplitude brightenings. A genuine outburst may be missed if only its low-amplitude onset or decay is present in the data, if it occurs during a gap, or if it is diluted by nearby sources. Conversely, false positives may be produced by residual instrumental trends, imperfect detrending near sector boundaries, scattered-light features, or short-lived excursions caused by imperfectly removed outliers. Pulsation beating and temporary amplitude enhancement can also resemble a low-amplitude, short-duration brightening if the time span considered is too narrow. The accepted candidates are therefore not considered as confirmed outbursts; we treat them as candidate stellar brightenings whose final physical classification requires follow-up observations.

These limitations are particularly relevant for the short durations measured for several of our outburst-like events. Each duration estimate is based on the interval over which the smoothed light curve remains above a local threshold and therefore traces the high-confidence part of the event rather than necessarily the full physical outburst profile. This conservative definition helps avoid overestimating events affected by residual trends, but it can shorten the measured durations compared with studies based on higher signal-to-noise light curves or different duration definitions.

We further note here that our duration estimates also strongly depend on the cadence, as demonstrated in Table~\ref{tab:outburst_candidates}. The same candidate outburst event is measured to be shorter in the 20 s cadence data than in the 120 s time series simply because the uncertainty of the individual data points is larger; therefore, the threshold of detection is higher.

The pixel-level diagnostics show that light curve morphology alone is not sufficient for classifying short TESS brightenings as stellar outbursts. Most of the selected events exhibit compact moving residuals, background variations, or aperture-grazing signals that are more consistent with moving-object contamination or related pixel-level systematics than with a brightening of the target star. This conclusion is consistent with the TESS flare survey of compact stars by \citet{2024ApJS..271...57X}, who found that flare-like signals in hot subdwarfs and white dwarfs are strongly affected by contamination from nearby objects or companions. The distinction is important because stellar flares generally show an impulsive rise followed by a faster, approximately exponential decay and are often associated with magnetically active cool companions or nearby contaminating stars, whereas ZZ~Ceti outbursts are usually interpreted as hour-scale brightenings of the white dwarf connected to nonlinear pulsation physics and often show a broader elevated-flux phase. Several of our identified events are short enough that, in morphology alone, they could resemble flare-like transients more than the multi-hour outbursts reported for the best-studied cool DAVs.

We therefore carried out the pixel-level check as a necessary but not sufficient test of stellar origin. The three targets do not show the clear moving compact residuals seen in the events rejected because of asteroid-contamination signatures. However, their TESS apertures are not free of crowding: the relevant CROWDSAP values are about 0.18--0.19 for TIC~219442838, about 0.14 for TIC~307908863, and about 0.44 for TIC~630426257 in the sectors with accepted events. These values mean that nearby sources can contribute a substantial fraction of the aperture flux. TESS target-pixel diagnostics can rule out obvious moving-object contamination in the inspected frames, but they cannot exclude all unresolved companions, close blends, or faint nearby flare stars. We therefore accept only TIC~219442838, TIC~307908863, and TIC~630426257 as candidate stellar-outburst objects, not as confirmed classical ZZ~Ceti outbursters.

Comparing the accepted candidates, listed in Table~\ref{tab:outburst_candidates}, with the sample presented by \citet[Table~1]{2017ASPC..509..303B}, we find that TIC 219442838, TIC 307908863, and TIC 630426257 have photometric effective temperatures above the coolest ZZ~Ceti stars in which outbursts have most commonly been reported. Table~\ref{tab:retained_outburst_teff} summarises the Gaia EDR3 photometric temperatures used in this work together with published temperature ranges collected from the Montreal White Dwarf Database \citep{2017ASPC..509....3D}. If these brightenings are confirmed as classical ZZ~Ceti outbursts, they may indicate that outburst-like behaviour can occur over a wider temperature range than previously assumed. At present, however, their short durations, the finite TESS pixel scale, and the non-negligible crowding in the relevant apertures require a more cautious interpretation. However, the spread of published values also shows that spectroscopic confirmation would be desirable.

\begin{table*}[!t]
\centering
\caption{Effective-temperature context for the accepted stellar-outburst candidates.}
\label{tab:retained_outburst_teff}
\begin{tabular}{rlcc}
\hline\hline
TIC ID & Common name & Gaia EDR3 $T_{\mathrm{eff}}$ & Published $T_{\mathrm{eff}}$ range \\
       &             & [K] & [K] \\
\hline
219442838 & EC 11266-2217 / WD 1126-222 & 11609 & 11187--12270 \\
307908863 & G 117-B15A / WD 0921+354 & 12042 & 11599--12600 \\
630426257 & GD 842 / WD 0120+002 & 12258 & 11517--12083 \\
\hline
\end{tabular}
\begin{flushleft}\footnotesize
Notes: The Gaia EDR3 values are the hydrogen-atmosphere photometric solutions used in Table~\ref{tab:outburst_candidates}. The published ranges list the span of literature values found in the Montreal White Dwarf Database.
\end{flushleft}
\end{table*}

The accepted candidate events are also shorter than those reported by \citet{2017ASPC..509..303B}. This difference may reflect intrinsic diversity among outbursting DAVs, but it may also be affected by the detection and measurement procedures, the smoothing timescale adopted here, the shorter or interrupted TESS observing windows, and flux dilution. It may also indicate that some accepted events are not classical ZZ~Ceti outbursts, but flare-like or blended stellar brightenings that are not separable with the available TESS data alone. Therefore, we treat the inferred durations as approximate estimates and emphasise that independent follow-up photometry and spectroscopic atmospheric parameters are required to confirm the stellar nature of these candidates.

We note that TIC 366681966, the white dwarf WD 1149+057, is already known as an outbursting ZZ Ceti star and was included in the sample of \citet{2017ASPC..509..303B}. It is also one of the two known outbursting DAVs with a splitting-based rotation estimate because \citet{2015ApJ...810L...5H} identified a stable triplet attributed to rotation. However, the specific TESS event identified by our search shows suspicious pixel-level behaviour in our diagnostics attributed to minor-planet crossing. We therefore exclude this event from the accepted stellar-outburst sample; however, this does not affect the previous classification of WD 1149+057 as an outbursting DAV or its literature status as an outbursting DAV with rotational splitting based on independent data.

To our knowledge, no previous outburst-like brightenings have been reported for TIC 219442838 (EC 11266-2217), TIC 307908863 (G 117-B15A), or TIC 630426257 (GD 842). If confirmed as originating from the white dwarfs rather than from blends or companions, these events would therefore represent new candidate ZZ~Ceti outbursts in these stars.

Among the accepted candidates, TIC 219442838 is the strongest case for recurrent stellar outburst-like behaviour. Four independent 120~s cadence brightenings are detected in different TESS sectors, and all show a similar short-duration morphology when plotted on the same relative time scale (Fig.~\ref{fig:tic219_outburst_overview}). This recurrence supports a stellar, non-asteroid interpretation for TIC 219442838, but it does not uniquely distinguish between a white dwarf outburst and another stellar transient source within or close to the TESS aperture. Independent follow-up observations would still be required for confirmation.

As an additional check for possible cool companions or blended red sources, we inspected archival broadband photometry from the TESS Input Catalogue \citep{2018AJ....156..102S} for the three accepted stellar-outburst candidates and for the 11 triplet rotators. The corresponding spectral-energy-distribution (SED) diagnostics are shown in Appendix~\ref{app:sed_diagnostics}. TIC~219442838 and TIC~307908863 have 2MASS near-infrared measurements and show no strong near-infrared excess in these data, while TIC~630426257 has no catalogued 2MASS or AllWISE counterpart in the TIC entry. Thus, the accepted outburst-candidate sample does not show clear evidence for cool companions in the available broadband photometry. Among the triplet rotators, TIC~313109945 shows the strongest red/infrared excess or blend signature ($G-K_{\rm s}\simeq5.2$ and $G-W1\simeq5.4$), consistent with the very low TESS CROWDSAP value of this target. We verified the WISE measurements against the AllWISE Source Catalogue and used only measurements with $S/N\geq3$. We find that TIC~394015496 shows a possible W2 excess, but the angular resolution of WISE means that blending remains a concern. We therefore flag TIC~313109945 and TIC~394015496 as possible cool-companions or blends that merit follow-up, but not as reliable companion detections.

\subsection{Rotation and outburst activity}

The accepted triplet sample and the identified stellar-outburst candidate sample from our homogeneous TESS analysis are disjoint. None of the 11 stars with accepted triplet-based rotation periods are among the three stars for which the pixel-level diagnostics support a candidate stellar outburst, and none of the three outburst candidates have an accepted rotational triplet in our extracted pulsation frequencies. This does not imply that the outburst candidates lack rotationally split modes; it only means that no complete triplet passed our conservative TESS detection and validation criteria for these stars. Thus, with the present TESS validation alone, we cannot measure a correlation between rotation period and outburst duration, amplitude, or recurrence time: the quantities are not simultaneously available for any star in the final validated TESS samples.

However, two such objects are already known from the literature. The prototype outbursting DAV KIC~4552982 has a rotationally split pulsation mode from which \citet{2015ApJ...809...14B} derived a probable rotation period of $17.47\pm0.04$~h. For PG~1149+057, \citet{2015ApJ...810L...5H} identified a stable triplet attributed to rotation and quoted a corresponding approximate rotation period of about 1.2~d, or about 29~h, while deferring a full period determination to future work. The first four confirmed Kepler/K2 outbursting DAVs were later summarised by \citet{2016ApJ...829...82B}, who noted that the first two cases showed short-period stable modes that appeared to be $\ell=1$ modes from rotational splittings.

We therefore distinguish two related but not identical questions. The present triplet analysis measures rotation through pulsation splittings and is most naturally interpreted as a mode-weighted seismological rotation rate, as opposed to a direct measurement of surface rotation. As the examples of PG~0112+104 and G226-29 show (see Sect.~\ref{sect:discussion-rotmultiplets}), surface-sensitive photometric signals can also be sought in pulsating white dwarfs, but their interpretation requires care. A test of a connection between outbursts and surface rotation would require reliable surface rotation periods for stars with reliably identified outbursts. We do not have such a matched sample here: G226-29 has an accepted triplet but no accepted stellar-outburst event, and its low-frequency signal does not provide a reliable surface-rotation period. The present TESS analysis, therefore, cannot directly relate outburst recurrence, duration, or amplitude to surface rotation. It only shows that our conservatively selected triplet rotators do not overlap with the conservatively accepted outburst candidates. This absence of overlap provides no new positive evidence for a rotation–outburst connection, but it also does not rule one out because the samples are small, the outburst candidates are not yet confirmed classical ZZ~Ceti outbursts, and the non-detection of a complete triplet does not imply absence of rotation.

The correct interpretation is therefore twofold. First, in the literature sample of known outbursting DAVs, there are already at least two stars with rotational splittings, KIC~4552982 and PG~1149+057. Second, if we restrict ourselves to the final samples selected homogeneously in this work, the 11 accepted TESS rotators and the three TESS outburst candidates have zero overlap. If we include the previously known outbursting DAV PG~1149+057, which is present in our TESS parent sample and has an approximate literature splitting-based rotation estimate, the comparison within the TESS parent sample becomes 12 stars with triplet- or splitting-based rotation information and four outbursting or candidate-outbursting stars, with one object in common. For 12 rotators and four outbursting or candidate-outbursting stars drawn independently from the 239 TESS-observed ZZ~Ceti targets, the expected random overlap is still only 0.20 stars. The probability of finding zero overlapping objects would be 0.813, and the probability of finding at most one overlapping object is 0.987. Thus, the present TESS data do not provide statistically significant evidence that outburst occurrence depends on the measured triplet or splitting-based rotation rate. Conversely, the two known outbursting rotators and the small expected overlap mean that the present data also cannot be used to argue against such a connection. The useful constraint from the present work is observational: within our conservative TESS triplet and outburst validation criteria, no new ZZ~Ceti star in our final validated TESS samples has both a secure triplet-based rotation period and is an accepted stellar-outburst candidate.

For the 11 accepted rotators from our TESS triplet search, the measured rotation periods span approximately 8--94~h, but all have zero accepted stellar-outburst events in our validated sample. For the three accepted outburst candidates, the event durations and peak amplitudes are listed in Table~\ref{tab:outburst_candidates}, but no triplet-based rotation period is available from this analysis. The broader literature context is therefore not secondary to the result, but part of it: KIC~4552982 and PG~1149+057 demonstrate that outbursting DAVs can be rotators with measurable splittings, whereas the present homogeneous TESS analysis does not add a further secure overlap object.

\section{Summary and conclusions}
\label{sect:summary}

We analysed TESS short- and ultrashort-cadence observations of a sample of known ZZ~Ceti stars, using all available data up to Sector 96. Our goal was to search for rotationally split pulsation multiplets suitable for estimating stellar rotation periods, to identify possible outburst-like brightenings in the corrected light curves, and to investigate whether any relation between the two phenomena is present. This question is motivated by the fact that two outbursting DAVs, KIC~4552982 and PG~1149+057, already have splitting-based rotation estimates in the literature.

After downloading and correcting the available TESS light curves, we performed frequency analysis and iterative prewhitening on the merged data sets. We then searched the resulting frequency lists with an automated consensus-clustering pipeline designed to identify and rank closely spaced, approximately equidistant triplet structures before final visual validation. Out of 417 analysed frequency lists, our automatic algorithm identified 44 triplet candidates in 30 TIC targets; 11 stars passed the final visual validation and were accepted as rotators. This corresponds to 4.6 percent of the 239 TESS-observed targets, but this value reflects the detection efficiency for clean, complete triplets rather than the intrinsic fraction of rotating ZZ~Ceti stars. In particular, noisy data can reduce an intrinsic triplet to an observed doublet, singlet, or non-detection under our selection criteria. Assuming first-order rotational splitting and adopting $C_{k,\ell}=0.47$, we derived rotation periods from the mean frequency separations. For objects with both 120 s and 20 s cadence detections, we averaged the splittings from the two data sets before calculating a single rotation period.

The resulting rotation periods span from about 8 h to 94 h. Most of the targets fall within the mass range typical of ZZ~Ceti stars, and our sample does not contain very high-mass white dwarfs. A comparison with the literature compilation of \citet{2017ApJS..232...23H} shows that our objects are broadly consistent with the large dispersion in rotation periods observed among normal-mass pulsating white dwarfs. The previously suggested tendency for the most massive white dwarfs to rotate faster is therefore not strongly tested by our sample.

We also searched the corrected TESS light curves for outburst-like brightenings. After automatic candidate selection and light curve inspection, 19 events in 11 stars passed to pixel-level validation. The target-pixel diagnostics show that most of these brightenings have signatures consistent with asteroid contamination or other suspicious features. We therefore accept only TIC 219442838, TIC 307908863, and TIC 630426257 as candidate stellar-outburst objects. Their measured events are short, their TESS apertures are affected by non-negligible crowding, and flare-like or blended stellar transients cannot be excluded with the available TESS data alone. These properties require confirmation with independent follow-up photometry and spectroscopic atmospheric parameters.

The 11 accepted TESS triplet rotators and the three identified TESS stellar-outburst candidates do not overlap, so the present homogeneous TESS analysis does not add a new secure outbursting rotator. This must be stated together with the literature fact that KIC~4552982 and PG~1149+057 are already known outbursting DAVs with rotational splittings, with reported periods of $17.47\pm0.04$~h and about 1.2~d, respectively. Including PG~1149+057 as a previously known outbursting DAV in our TESS parent sample gives one overlap among 12 stars with triplet- or splitting-based rotation information and four outbursting or candidate-outbursting stars. This remains statistically unsurprising for a parent sample of 239 objects. We therefore find no new evidence in the present TESS data for a measurable relation between triplet-based rotation periods and outburst activity, but the existing two outbursting rotators and the small samples also prevent us from excluding such a relation. Although spot-like modulation or other non-pulsational photometric variability has been reported in individual pulsating white dwarfs, including our triplet target G226-29, the origin of the variable signal in G226-29 is uncertain, and the three accepted TESS outburst candidates have no reliable surface-rotation periods. Thus, we are not in the position to directly test a possible connection between outbursts and surface rotation.

In summary, our analysis expands the set of TESS-observed ZZ~Ceti stars with candidate rotationally split triplets and provides new rotation-period estimates for 11 objects. In addition, we identify three stellar outburst candidates after rejecting several light-curve-selected events with suspicious pixel-level behaviour.
The results demonstrate the continued value of extended TESS coverage for studying both the internal rotation and the nonlinear photometric behaviour of pulsating white dwarfs, while also showing that pixel-level validation is essential for short outburst-like TESS events.

\section*{Data availability}
Table~\ref{tab:mpchecker_matches} is only available in electronic form at the CDS via anonymous ftp to \url{cdsarc.u-strasbg.fr} (130.79.128.5) or via \url{http://cdsweb.u-strasbg.fr/cgi-bin/qcat?J/A+A/}.

\begin{acknowledgements}

We thank the anonymous referee for the careful reading of the manuscript and for the constructive comments.

The authors thank Keaton Bell (CUNY Queens College, Department of Physics) and Weikai Zong (Beijing Normal University) for their helpful comments on the manuscript.

The authors also acknowledge the financial support of the KKP-137523 `SeismoLab' \'Elvonal grant from the Hungarian Research, Development and Innovation Office (NKFIH).

This paper includes data collected with the TESS mission, obtained from the MAST data archive at the Space Telescope Science Institute (STScI). Funding for the TESS mission is provided by the National Aeronautics and Space Administration (NASA) Explorer Programme. STScI is operated by the Association of Universities for Research in Astronomy, Inc., under NASA contract NAS 5–26555.

\end{acknowledgements}

%-------------------------------------------------------------------

%\begin{thebibliography}{}
%\end{thebibliography}

\bibliographystyle{aa} % style aa.bst
\bibliography{rotation} % your references Yourfile.bib

\begin{appendix}
\onecolumn
\section{Broadband SED diagnostics}
\label{app:sed_diagnostics}
\begin{figure}[!ht]
\centering
\includegraphics[width=\textwidth,height=0.76\textheight,keepaspectratio]{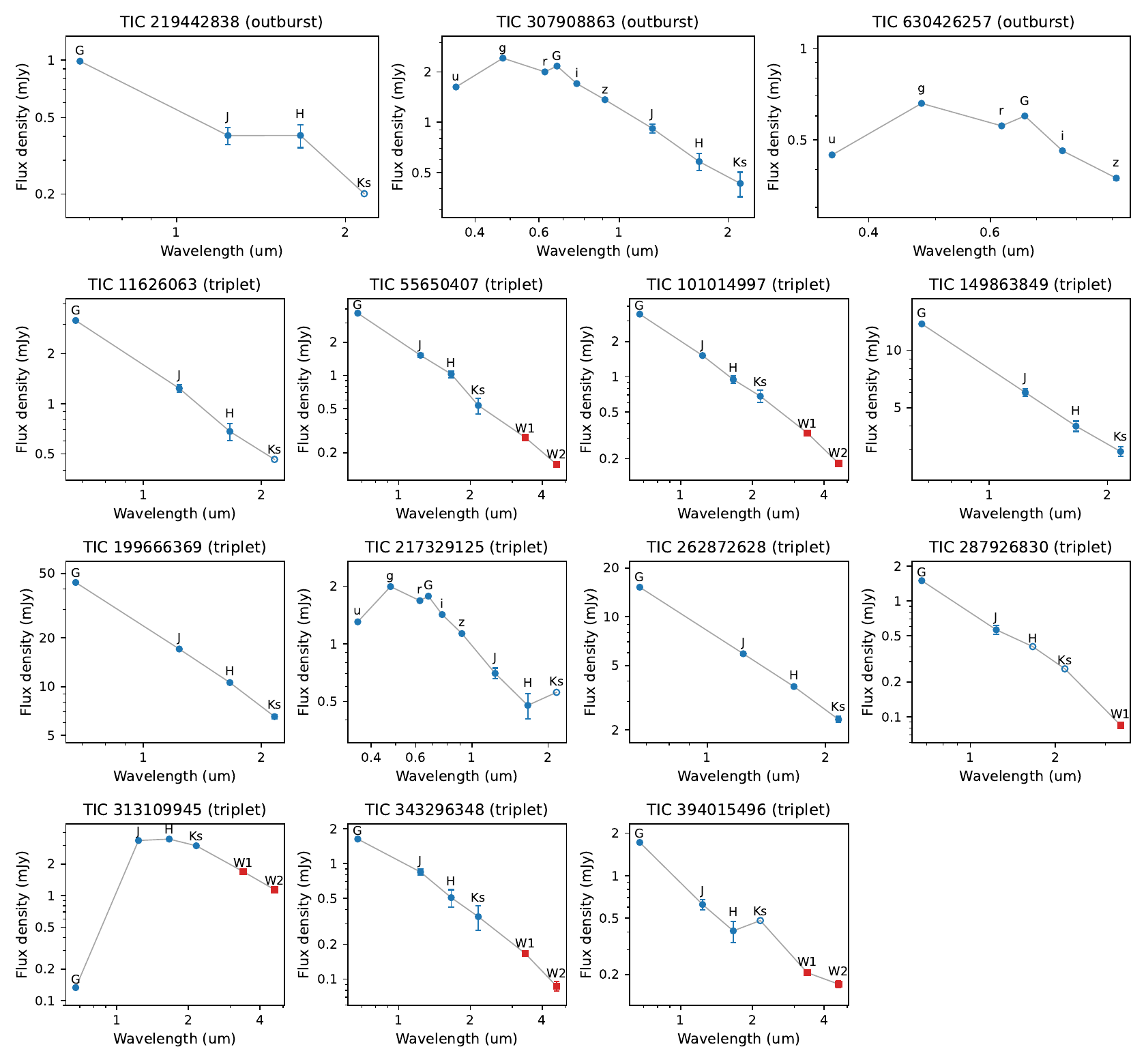}
\caption{Broadband SED diagnostics for the three accepted stellar-outburst candidates and the 11 accepted triplet rotators, based on photometry retrieved from the TESS Input Catalog. Blue points show Gaia/SDSS/2MASS bands, and red squares show AllWISE bands where available. Vertical error bars represent catalogue magnitude uncertainties converted to flux-density uncertainties; open symbols indicate non-WISE bands for which no uncertainty is listed, and error bars smaller than a symbol may be visually indistinguishable. Only WISE measurements with $S/N\geq3$ are plotted; all W3 and W4 entries failed this quality criterion and were excluded. The accepted outburst candidates do not show clear evidence for a cool companion in the available TIC photometry.}
\label{fig:app_sed_diagnostics}
\end{figure}

\clearpage

\section{Representative pixel-level diagnostics of events with asteroid-contamination signatures}
\label{app:asteroid_pixel_diagnostics}
\begin{figure}[!ht]
\centering
\includegraphics[width=\textwidth,height=0.58\textheight,keepaspectratio]{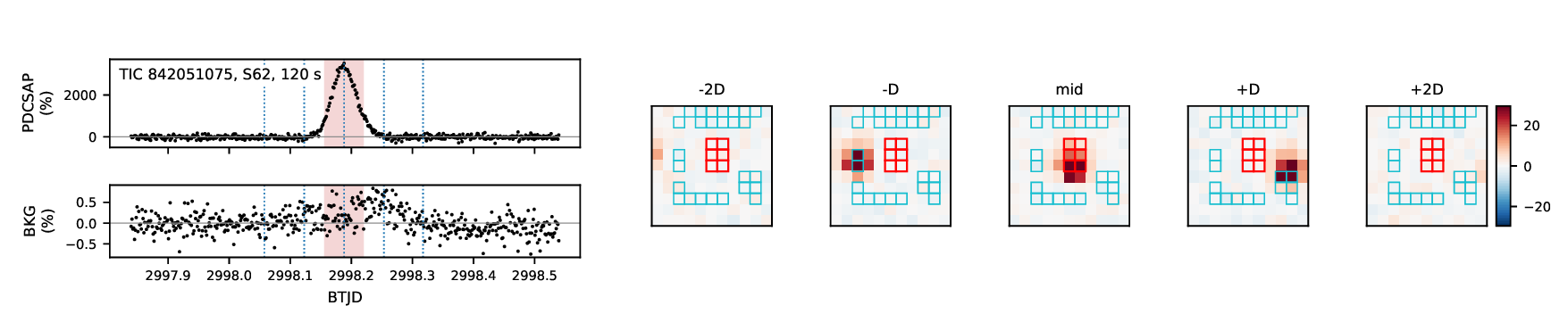}
\caption{Representative pixel-level diagnostics for an outburst-like brightening that shows signatures consistent with asteroid contamination or other suspicious features in the TESS target-pixel data. This event was omitted from the candidate stellar-outburst sample. The two panels on the left show the SPOC PDCSAP light curve and the corresponding SAP background level around the event. The red shaded interval marks the measured brightening, and the vertical dotted lines mark the five Target Pixel File (TPF) frames shown on the right. These frames sample the event midpoint and times separated from it by one and two measured event durations. The five image panels show difference images, computed as individual TPF frames minus the local median baseline image. Red contours mark the SPOC photometric aperture, while cyan contours mark the background pixels. A compact, moving positive residual outside or grazing the aperture, especially when accompanied by changes in the background level, indicates that the apparent brightening is likely caused by a moving Solar System object or by aperture/background contamination rather than by a stellar outburst. Only this representative example is shown here to illustrate the rejection criterion; the full set of rejected-event diagnostics is provided in the electronic material.}
\label{fig:app_asteroid_pixels1}
\end{figure}

\section{Pixel-level diagnostics of the accepted candidate stellar outbursts}
\label{app:outburst_pixel_diagnostics}
\begin{figure}[!ht]
\centering
\includegraphics[width=\textwidth,height=0.58\textheight,keepaspectratio,trim=0 130 0 0,clip]{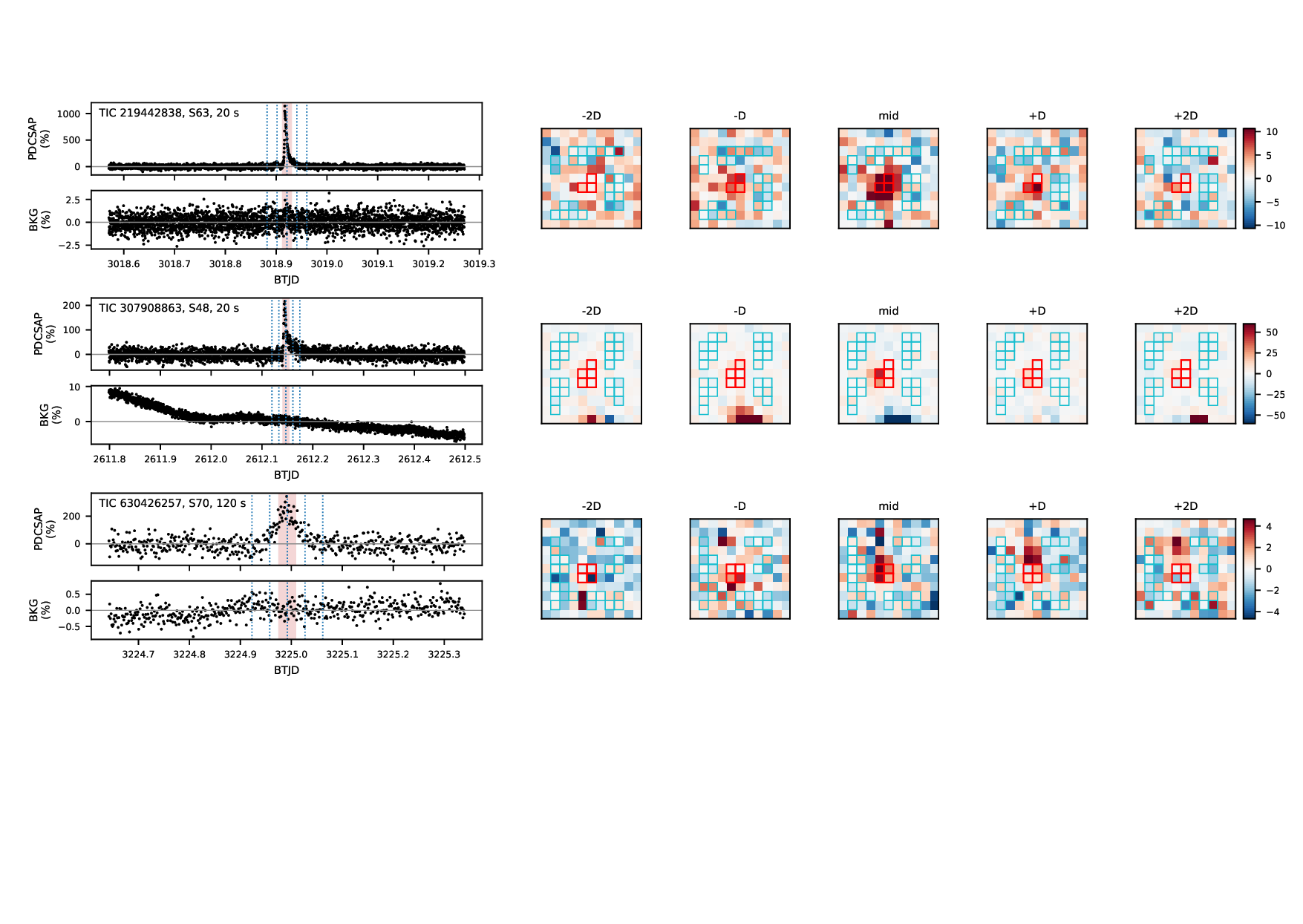}
\caption{Pixel-level diagnostics for the accepted candidate stellar outbursts. The panels are arranged as in Appendix~\ref{app:asteroid_pixel_diagnostics}. In contrast to the representative cases with asteroid-contamination signatures shown in Fig.~\ref{fig:app_asteroid_pixels1}, these events do not show a clear moving compact residual crossing the aperture or background pixels in the inspected TPF frames. They are therefore considered as candidate stellar brightenings, although the available TESS data do not by themselves prove that they are classical ZZ~Ceti outbursts from the target white dwarfs.}
\label{fig:app_outburst_pixels}
\end{figure}

\clearpage

\section{Electronic material}
\label{app:electronic_material}

\begin{figure}[!ht]
\centering
\includegraphics[width=0.84\textwidth,height=0.66\textheight,keepaspectratio]{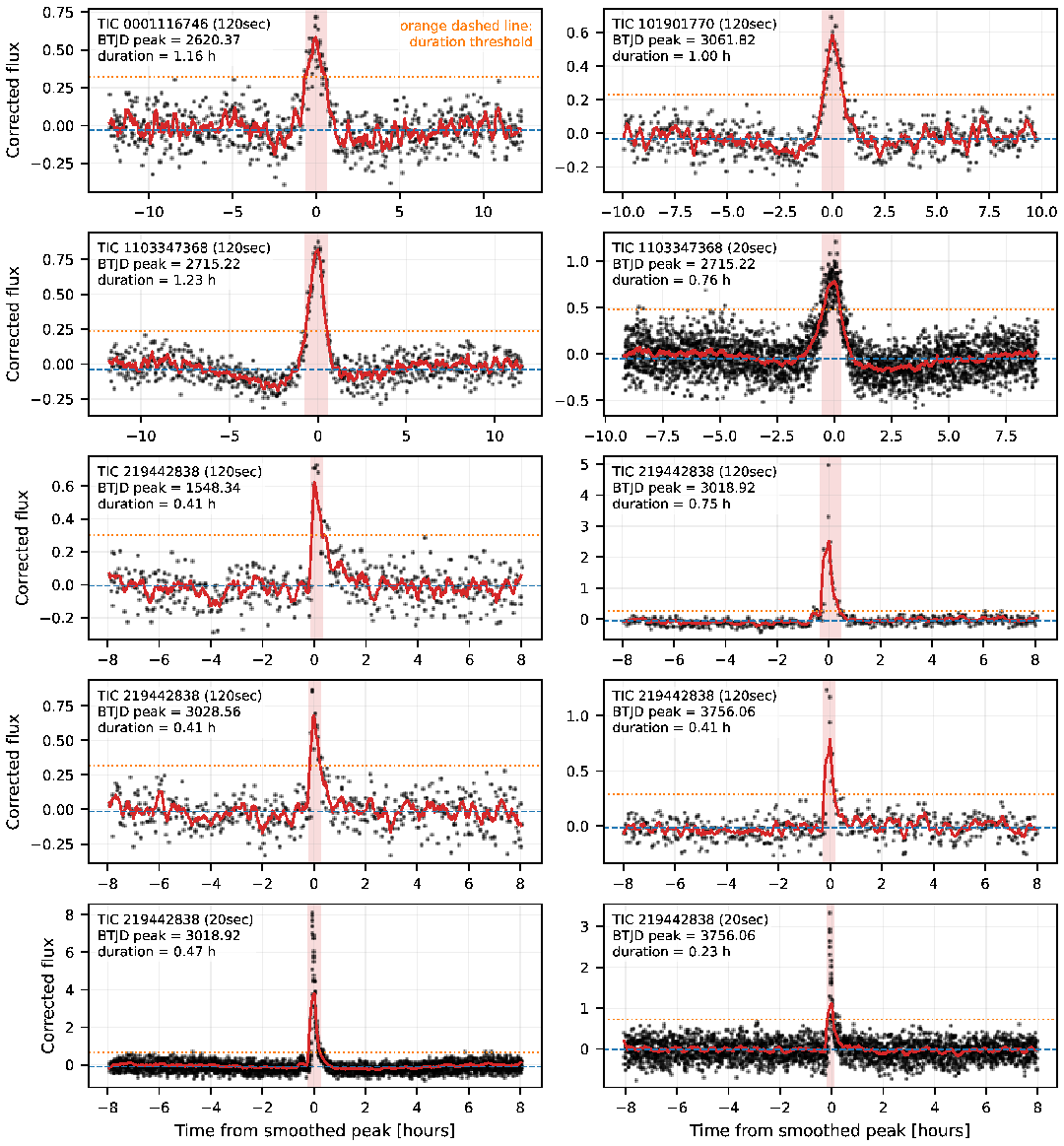}
\caption{Outburst-like brightenings detected in the corrected TESS light curves of ZZ~Ceti stars. Each panel shows the flux variation around an individual candidate event, centred on the peak of the smoothed light curve. Grey points denote the corrected TESS fluxes, while the black curve shows the smoothed light curve used to estimate the event duration. The horizontal dashed line marks the duration threshold relative to the local quiescent flux level. The measured duration and the BJD time of maximum brightness are indicated in each panel. These events are shown before pixel-level validation.}
\label{fig:electronic_outburst1}
\end{figure}

\clearpage

\begin{figure}[!ht]
\centering
\includegraphics[width=0.84\textwidth,height=0.76\textheight,keepaspectratio]{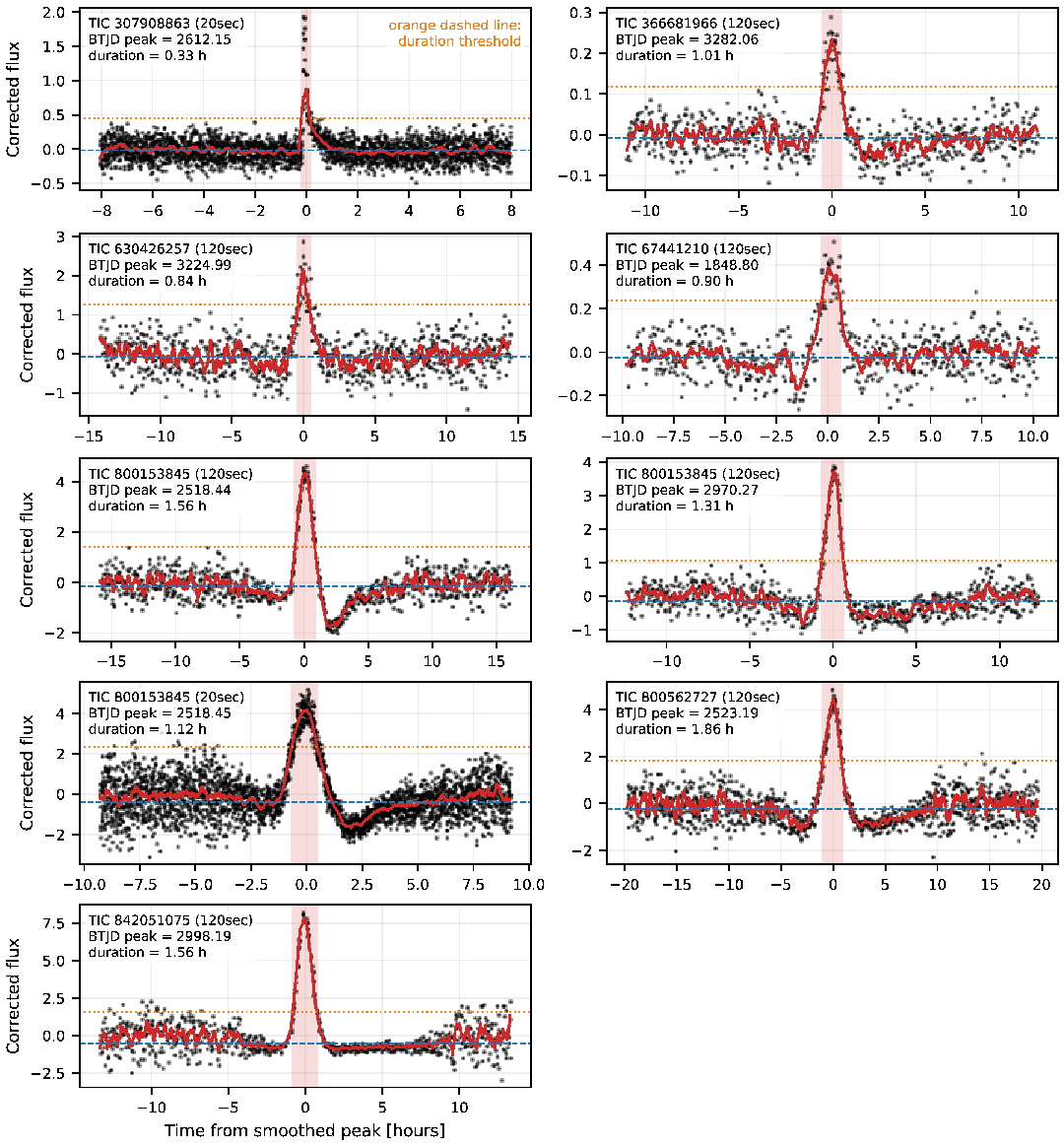}
\caption{Continuation of Fig.~\ref{fig:electronic_outburst1} for the remaining light-curve-selected outburst-like events. The panels show the corrected TESS fluxes around the selected brightening events, together with the smoothed light curves and the adopted duration thresholds used to estimate the event durations. These are light-curve-selected events and are not all accepted as candidate stellar outbursts.}
\label{fig:electronic_outburst2}
\end{figure}

\clearpage

\begin{figure}[!ht]
\centering
\includegraphics[width=\textwidth,height=0.76\textheight,keepaspectratio]{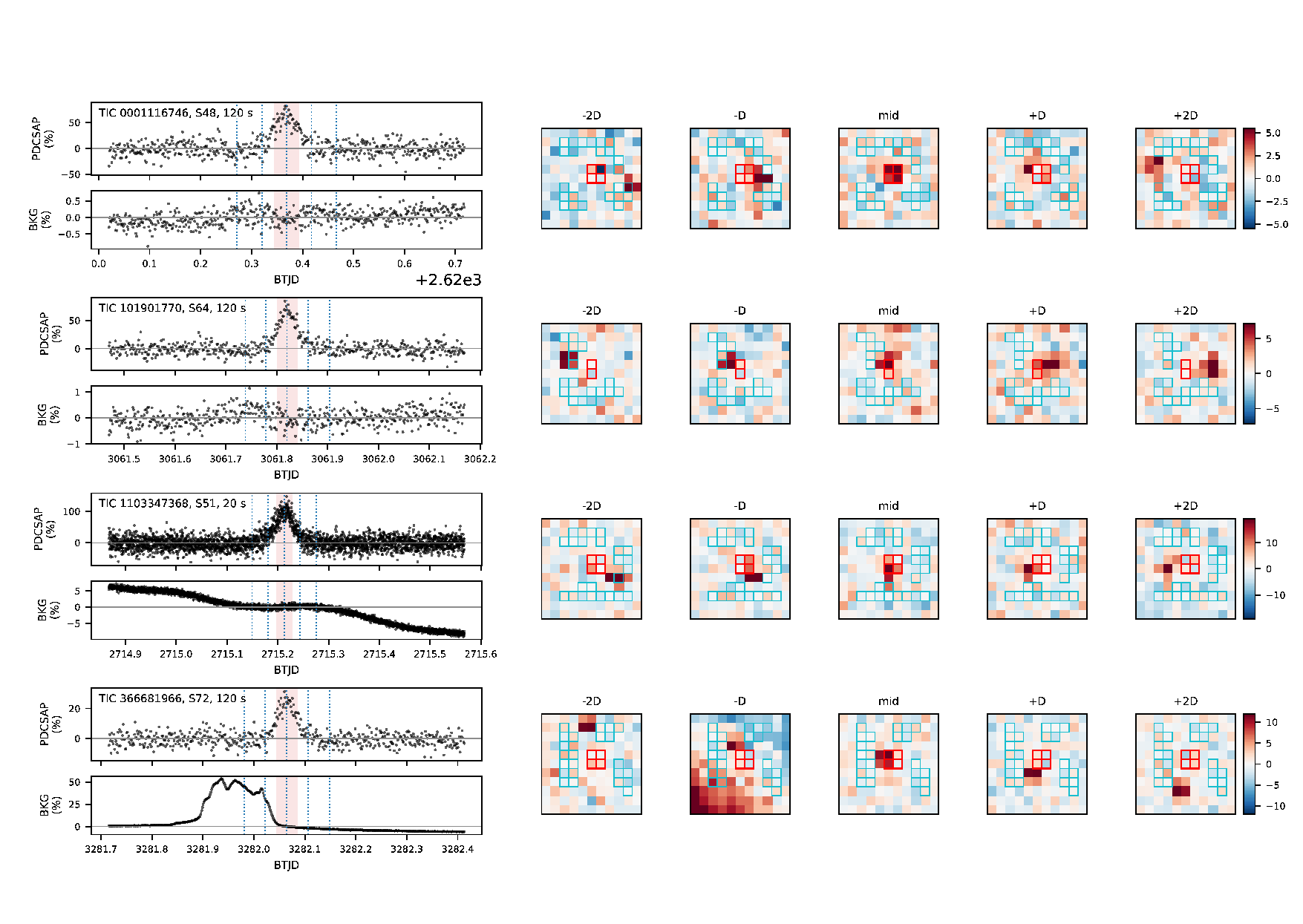}
\caption{Pixel-level diagnostics for outburst-like brightenings that show signatures consistent with asteroid contamination or other suspicious behaviour in the TESS target-pixel data. Each row corresponds to one event. The two panels on the left show the SPOC PDCSAP light curve and the corresponding SAP background level, while the five panels on the right show event-centred difference images. Red contours mark the SPOC photometric aperture, and cyan contours mark the background pixels. These events were omitted from the candidate stellar-outburst sample.}
\label{fig:electronic_asteroid1}
\end{figure}

\clearpage

\begin{figure}[!ht]
\centering
\includegraphics[width=\textwidth,height=0.76\textheight,keepaspectratio]{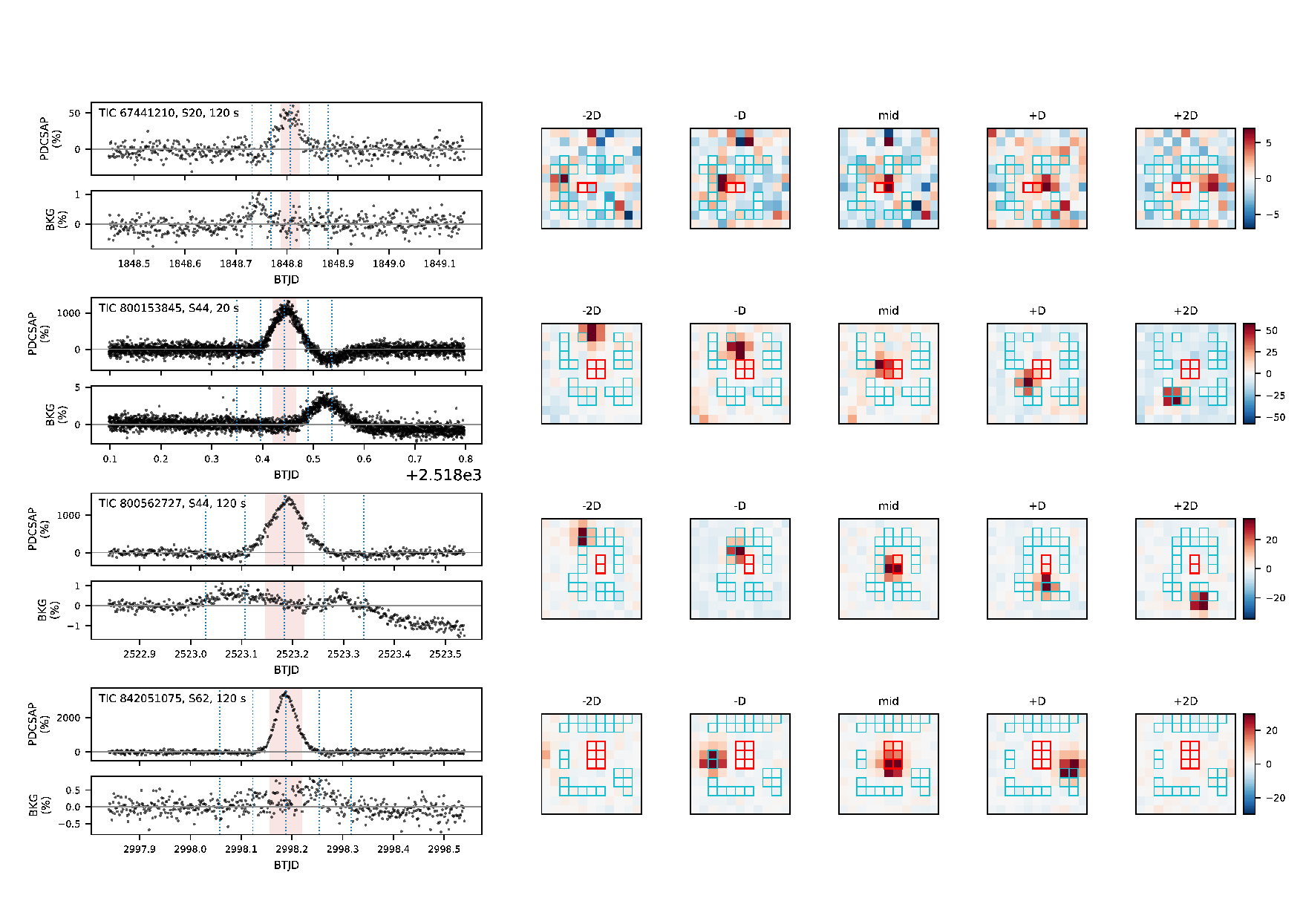}
\caption{Continuation of Fig.~\ref{fig:electronic_asteroid1}. The same diagnostics are shown for additional outburst-like brightenings with signatures of asteroid contamination or other suspicious pixel-level behaviour.}
\label{fig:electronic_asteroid2}
\end{figure}

\end{appendix}

\end{document}